\documentclass[sigplan,10pt,nonacm]{acmart}
\renewcommand\footnotetextcopyrightpermission[1]{}

\setcopyright{none}

\usepackage{iftex}
\ifPDFTeX
  \usepackage[utf8]{inputenc}
\fi

\usepackage{listings}
\usepackage{amsmath}
\usepackage{pifont}
\usepackage{booktabs}
\usepackage{array}
\usepackage{calc}
\usepackage{pgfplots}
\pgfplotsset{compat=1.17}

\usepackage{iftex}
\ifPDFTeX
  \providecommand{\unicodemono}{\ttfamily}
\else
  \usepackage{fontspec}
  \IfFontExistsTF{DejaVu Sans Mono}
    {\newfontfamily{\unicodemono}{DejaVu Sans Mono}[Scale=0.85]}
    {\providecommand{\unicodemono}{\ttfamily}}
\fi

\providecommand{\passthrough}[1]{#1}

\newcommand{\sysfull}{CoAgent}
\newcommand{\sys}{CoAgent}
\newcommand{\mtpo}{MTPO}
\newcommand{\cmark}{\ding{51}}
\newcommand{\xmark}{\ding{55}}

\newcommand{\phead}[1]{\par\indent\textbf{\textit{#1}}\enspace\ignorespaces}

\title{\sysfull{}: Concurrency Control for Multi-Agent Systems}

\author{Hongtao Lyu}
\affiliation{
  \institution{Shanghai Jiao Tong University}
  \country{China}}
\email{hongtaolyu@sjtu.edu.cn}

\author{Dingyan Zhang}
\affiliation{
  \institution{Shanghai Jiao Tong University}
  \country{China}}
\email{healthcliff-ding@sjtu.edu.cn}

\author{Mingyu Wu}
\affiliation{
  \institution{Shanghai Jiao Tong University}
  \country{China}}
\email{mingyuwu@sjtu.edu.cn}

\author{Xingda Wei}
\affiliation{
  \institution{Shanghai Jiao Tong University}
  \country{China}}
\email{wxdwfc@sjtu.edu.cn}

\author{Haibo Chen}
\affiliation{
  \institution{Shanghai Jiao Tong University}
  \country{China}}
\email{haibochen@sjtu.edu.cn}

\begin{abstract}
Multi-agent LLM systems---coding agents, devops agents, document agents---now routinely run several agents in parallel against the same git tree, Kubernetes cluster, or document.
As soon as two of them mutate shared state, they enter the regime classical concurrency control has studied for decades, but classical mechanisms fit LLM agents poorly.
A single agent transaction spans minutes of inference, read sets are broad and opaque rather than statically inferable, and the live state agents act on admits neither fork nor buffer, so writes take effect the moment they execute.
Locks block long inference intervals; OCC abort-and-retry discards minutes of work on every conflict.

This paper builds concurrency control on a capability classical transactions lack: the LLM inside each agent can judge whether a conflicting write actually invalidates its plan, and can repair exactly the operations that depended on it.
Control therefore turns advisory: the runtime informs, the agent repairs.
Our protocol, \mtpo{} (Monotonic Trajectory Pre-Order), fixes a serialization order at launch, serves each read the order-filtered value, and applies writes speculatively in place; a one-way notification asks an affected reader to re-judge and patch its plan, while the framework mechanically undoes and reorders misplaced writes through the saga-style inverse each tool registers in advance.
At quiescence the run is serializable in the pre-decided order.
We realize \mtpo{} as \sysfull{}, toolcall middleware whose privileged ToolSmith grows footprint-declared, undoable tools online.
On ten contended workloads, \sysfull{} stays within 5\% of serial correctness at a 1.4$\times$ speedup and near-serial token cost, where 2PL and OCC surrender nearly all concurrency gains; on a \texttt{bash}-only target system, it grows a 25-tool library online and lifts the task pass rate from 45/71 to 63/71 at 0.80$\times$ the time and 0.86$\times$ the cost.
\end{abstract}

\begin{document}

\maketitle
\pagestyle{plain}

\section{Introduction}
\label{introduction}

\noindent
LLM-powered agents, which take actions based on LLM outputs,
have revolutionized the artificial intelligence landscape,
enabling end-to-end goal achievement on tasks like writing and debugging code~\cite{anthropicclaude2025,carlini2026compiler},
research automation~\cite{karpathy2026autoresearch},
and deep research~\cite{hadfield2025multiagent}, just to name a few~\cite{DBLP:conf/iclr/YaoZYDSN023}.
A single agent, however, can be inefficient due to sequential generation of LLM outputs
and incapable of complex tasks and
may hit a structural ceiling due to context limitations~\cite{kimi2026agentswarm}. 
Modern systems therefore decompose the task
and run multiple sub-agents concurrently,
and the multi-agent paradigm has accordingly become
a key pillar of modern agentic systems:
Claude Code ships parallel sub-agent teams~\cite{anthropicclaude2025,claudecodeagents2024,carlini2026compiler},
OpenAI Codex and Cursor run background agents on parallel tasks~\cite{openai2025codex,cursor2025worktrees},
and Kimi Agent Swarm coordinates up to a hundred agents on one job~\cite{kimi2026agentswarm}.

Running multiple sub-agents concurrently revives a classical problem.
Each sub-agent action reads or writes shared state---a
file on disk, a page on the web, the configuration of a Kubernetes (K8S) cluster. 
Thus, like traditional concurrent programs,
the interleaving of reads and writes can easily lead to anomalies. 
{Such failures are already documented at scale: a recent audit of over
200 multi-agent traces attributes more than a third of all failures to
inter-agent misalignment, including agents acting on the same resource unaware
of each other~\cite{cemri2025mast}.}
Consider a concrete instance that we measured on a live K8S cluster,
where two agents
operate concurrently.
Agent A handles a remediation task from AIOpsLab~\cite{aiopslab2025}:
a faulty rollout has left several deployments running a wrong container image,
and A must find and repair every affected deployment.
Agent B prepares a canary release for one service:
it reads the service's current image and creates a canary deployment mirroring that image.
In this run, A scans the cluster before the canary exists, so its repair misses the canary;
B reads the image before A repairs it, so the canary is built on the bad image.
Both agents faithfully complete their tasks and report success,
yet the cluster ends with a canary on the known-bad image, silently waiting to take live traffic.
Neither agent erred in its own reasoning; the fault lies in the
interleaving of their reads and writes, and the final cluster state matches no
serial execution of the two agents.
\textsection{\ref{multi-agent-consistency-problem}} walks through the trace step by step.

This paper studies how to provide concurrency control for multi-agent systems.
Current systems (\textsection{\ref{existing-agent-concurrency-practice}}) 
either run one agent at a time~\cite{wu2023autogen,qian2024chatdev, hong2024metagpt, crewai2024},
statically partition the agents' write sets so that no two agents touch the same state~\cite{claudecodeagents2024},
or fork the shared state per agent and merge the forks once every agent finishes \cite{cursor2025worktrees,cognition2024devin,replitagent4}.
None of these is ideal.
Running one agent at a time loses the concurrency benefits of multiple agents;
static write partitioning requires the write set to be known in advance,
which is often impossible for agents whose plans emerge during execution;
and fork-and-merge offers only weak isolation akin to read committed in databases~\cite{berenson1995critique},
so it cannot prevent anomalies like the one described above (detailed in {\textsection{\ref{multi-agent-consistency-problem}}}).

The ideal property to enforce on a multi-agent system is therefore serializability~\cite{berenson1995critique},
which guarantees that any concurrent execution is equivalent to some serial order of the agents.
Unfortunately, the two classical protocols that realize it---two-phase locking (2PL) and optimistic concurrency control (OCC)---both 
fall short in the agent setting, 
due to two agent-related gaps (\textsection{\ref{goal-and-gap}}):

The first is a \emph{functionality gap}.
Classical protocols assume the shared state is plain data such as database rows---data that the system can read, buffer, and overwrite at will.
Agents, however, also act on \emph{logical state} whose effects exist only in the external world:
a \texttt{kubectl apply} in our K8S example must be executed against the cluster to take effect,
and the agent must observe that effect to continue debugging.
This breaks OCC's staging discipline,
which requires modifications to remain in a private buffer until commit.
2PL hits the same wall when it resolves a deadlock:
the victim's writes have already taken effect on the external world,
and the system has no way to roll them back ({\textsection{\ref{functionality-gap}}}).

The second is a \emph{performance gap}.
Classical protocols were designed for transactions that finish quickly.
Agents, in contrast, may run for hours, or even days~\cite{carlini2026compiler}.
At this timescale,
2PL forces an agent to hold its locks for the entire task,
blocking every other agent that touches overlapping state.
OCC instead pays at re-execution time:
even setting the functionality gap aside,
a single conflict at validation discards the agent's entire work.

To close these gaps,
our key insight is that the backbone of an agent---the LLM---has semantic understanding of the task and the state it is operating on,
and that understanding supplies the remedies classical CC lacks.
First, the LLM can separate \emph{real} conflicts from semantically irrelevant read--write interference:
most overlaps on an agent's broad read set invalidate no premise its plan relies on
(a peer appending a log line to a file the agent scanned changes nothing it depends on),
and the agent can dismiss them with no work lost.
Second, on a real conflict, the LLM can choose the most efficient repair path
in place of OCC's all-or-nothing restart:
it inspects its own read/write history,
identifies the few actions premised on the now-stale value,
and re-executes only those.
For the conflicts such repair cannot dissolve---a write already applied to the external world but now in the wrong order---the framework
falls back to saga-style compensation~\cite{sagas1987}:
every write registers an inverse before executing
(a \texttt{kubectl apply} is reversed by re-applying the manifest it displaced),
so the framework mechanically undoes and reorders misplaced writes without consulting the LLM.

Based on this insight, we propose \sysfull{} ({\sys}),
an optimistic concurrency-control framework for multi-agent systems.
{\sys} builds on optimism to unleash concurrent execution:
agents execute without being blocked by concurrent agents. 
To detect and fix anomalies during execution,
we propose a notification mechanism that
notifies each agent of concurrent actions that may break its isolation.
The notified agent judges whether the change invalidates its premises
and patches only the affected actions;
writes that landed in the wrong order are undone and reapplied by the framework itself,
using the inverses those writes registered.

{\sys} guarantees serializability as long as agents follow the protocol's mechanical rules and,
when notified, correctly judge which of their premises and pending actions the conflicting write affects;
\textsection{\ref{system-assumptions}} states both as explicit assumptions of our formal model.
These tasks are simple for frontier LLMs:
even on a budget-friendly model (DeepSeek v4 flash),
our evaluation observed a notified agent misjudging a notification's relevance in only 5\% of trials (\textsection{\ref{evaluation}}).
By keeping execution concurrent and recovery selective,
{\sys} also significantly improves both efficiency and functionality on multi-agent workloads
compared with existing concurrency control approaches: {it passes all ten
contended workloads with correctness within 5\% of serial execution, at a
1.4$\times$ speedup and near-serial token cost (1.15$\times$).
Under the same contention, 2PL deadlocks 0.81 times per trial and recovers
almost no speedup (1.04$\times$), while OCC aborts 0.95 times per trial, runs
slower than serial (0.93$\times$), and pays 1.83$\times$ token cost;
uncoordinated execution is fast but passes only 13\% of trials.
Beyond the contended workloads, on a target system exposing only \texttt{bash}
and no tool table, {\sys} grows a 25-tool library online and lifts the
task pass rate from 45/71 to 63/71, at 0.80$\times$ the time and
0.86$\times$ the cost of the \texttt{bash} baseline
(\textsection{\ref{sec:eval-toolgrowth}})}.

\section{Background \& Motivation}\label{background-motivation}

\subsection{Background: LLM-powered Multi-Agent Systems} 
\label{the-agent-paradigm-we-target}

The classical literature defines an agent as a software entity that perceives an environment through
observations, deliberates over those observations against some goal, and acts back on the
environment~\cite{wooldridge2009introduction,russell2021artificial}. Modern LLM-powered agent
systems instantiate this abstraction by using a large language model (LLM) as the deliberation
engine and explicit tool calls for observations and actions. This long-running, tool-centric
architecture has emerged as the dominant implementation pattern for contemporary coding and workflow
agents \cite{DBLP:conf/iclr/YaoZYDSN023,yang2024sweagent,anthropicclaude2025}. The execution model of such agents is characterized
by three key properties:

\textbf{Append-only context.} Each agent maintains a private \emph{context} for the current
session---system prompts, user instructions, tool calls, execution results, and model outputs.
The context is designed as an append-only structure, where
every LLM inference step appends new contents to the existing history. This structural design
effectively leverages optimizations such as prefix KV caching \cite{kwon2023pagedattention,zheng2024sglang},
significantly improving LLM inference efficiency.

\textbf{Shared external state and tool-call boundary.} Beyond the private context, the system
involves external states that can be shared across multiple agents (e.g., files, database tables,
Kubernetes objects, collaborative documents, and external API endpoints)---in practice, named
resources mutated through a small verb set, the interface REST canonizes~\cite{rfc7231}. Agents
rely on LLM inference to generate tool calls---the sole channel through which an agent observes
or mutates these states.

We distinguish two kinds of tool call. A \emph{read} yields an \emph{immutable perception}:
once returned, the value is recorded in the agent's context as a fixed premise. A \emph{write}
mutates the external state in one of two ways. A \emph{blind write} sets the state to a value
independent of what it replaces: overwriting a configuration key, deleting an entry (REST's
\texttt{PUT} and \texttt{DELETE}). A \emph{read--modify--write} (RMW) produces an effect that
composes with the prior state: appending to a collection, creating an entry under a
server-assigned id (REST's \texttt{POST}). Idempotence is the observable criterion that
separates the two~\cite{rfc7231}: replaying a blind write is harmless; replaying an RMW write
is not---two \texttt{POST}s create two entries.\footnote{Idempotent RMW writes exist---a
merge-style partial update (\texttt{PATCH}) re-applies cleanly---and we conservatively treat
them as RMW.} The difference is invisible to a single agent; under concurrency it decides
whether a misordered write can be repaired after the fact (\S\ref{detailed-design}).

\textbf{Agent view.} The agent's accumulated perceptions form its \emph{view} of the external
world---the sole basis for any subsequent reasoning or tool call. Once a read returns, its value is
recorded in the context unchanged: an agent that reads state \passthrough{\lstinline!X!} and obtains
\passthrough{\lstinline!v0!} will treat \passthrough{\lstinline!v0!} as the value of
\passthrough{\lstinline!X!} on every later inference, even after \passthrough{\lstinline!X!} has
been mutated externally to \passthrough{\lstinline!v1!}. The runtime must either trigger a re-read or push a change message for
the view to catch up; without one of those, every action premised on the now-stale
\passthrough{\lstinline!v0!} may be wrong. Closing this stale-view gap is the central concern of
this paper.

\subsection{The Multi-Agent Consistency Problem}
\label{multi-agent-consistency-problem}
To improve system efficiency and throughput, it is common to adopt a multi-agent architecture that
leverages multiple agents to process tasks concurrently. However, multi-agent concurrency issues
arise whenever multiple agents operate over the same shared state.

Figure~\ref{fig:canary-trace} shows how two agents, operating concurrently on a live Kubernetes
cluster, produce a textbook concurrency anomaly despite neither making an individual reasoning
error; the trace is a measured run of the case study that \S\ref{sec:eval-casestudy} replays under
four coordination protocols. Agent~A is given a remediation task taken verbatim from AIOpsLab: a
rollout has left an unknown subset of deployments on a wrong image, and A must locate every
affected deployment and restore its image to the canonical one (three here: \texttt{geo},
\texttt{profile}, \texttt{reservation}). Agent~B is independently tasked with preparing a canary
release for \texttt{geo}: read \texttt{geo}'s current image, then create a zero-replica deployment
\texttt{geo-canary} mirroring that image, to take a small share of live traffic at the next
release window. The anomaly arises due to interleaved reads. Agent~A lists the cluster before the
canary exists, so its audit scope is fixed at the three mismatched services; Agent~B reads
\texttt{geo}'s image before A repairs it, so the canary is built on the bad image. Both agents
then finish and report success: B has mirrored what it read, and A's closing check even lists the
new canary but judges it outside the incident it was asked to fix. Yet the final state is one no
serial order produces: run A first and the canary mirrors the repaired image; run B first and A's
sweep catches the canary as a fourth mismatch; either way the canary ends on the canonical image.
The failure is also silent: the zero-replica canary raises no alarm until the release window
routes live traffic onto the known-bad image.
The trace also exercises both write classes of \S\ref{the-agent-paradigm-we-target}:
Agent~A's image restorations are blind overwrites, while Agent~B's canary creation is a
read--modify--write; the distinction returns in \S\ref{detailed-design}.

\begin{figure*}[!t]
\begin{lstlisting}[columns=fixed,basicstyle=\unicodemono\scriptsize]
 t(s)  Action                        Shared-state change   A's view               B's view
 ----  ----------------------------  --------------------  ---------------------  -----------------------
 0     (faulty rollout)              geo.image = bad       ---                    ---
 3.9   A: list deployments                                 scope = {geo, +2}
 4.5   B: read geo.image                                                          geo.image = bad
 6.1   B: create geo-canary          canary.image := bad   scope (stale): misses  canary.image = bad
       mirroring the image it read                         the new canary
 8.2   A: fix the three mismatches   geo.image := good,+2                         geo.image = bad (stale)
 ----  ----------------------------  --------------------  ---------------------  -----------------------
\end{lstlisting}
\caption{\small{The canary anomaly, measured on an AIOpsLab-derived workload and replayed
in the case study of \S\ref{sec:eval-casestudy}: both agents faithfully execute their
assigned tasks, yet their interleaved reads and writes leave a wrong final state.}}
\label{fig:canary-trace}
\end{figure*}

\subsection{Existing Isolation Mechanisms for Multi-Agent}
\label{existing-agent-concurrency-practice}

Existing agent systems handle these consistency issues through three broad strategies, all with
significant drawbacks.

\textbf{Sequential execution.} The simplest strategy prevents stale views by prohibiting concurrent
execution entirely; during agent handoffs, consistency is preserved either by inheriting the fully
updated context~\cite{wu2023autogen} or by clearing the context to force a fresh state
re-read~\cite{qian2024chatdev, hong2024metagpt, crewai2024}. This is safe but falls back to serial
execution: multiple agents serve only to improve task success rates, not to reduce job completion
time.

\textbf{Write partitioning.} A second strategy statically partitions the agents' write sets,
ensuring no two agents concurrently modify the same object. In coding scenarios, systems like Claude
Code Agent Teams~\cite{claudecodeagents2024} explicitly prompt agents to operate on mutually
exclusive file sets. The canary trace already obeys this discipline (A repairs existing
deployments, B only creates new objects), yet the anomaly stands: disjoint write sets address
neither of the two relevant ACID properties. \textbf{Isolation (I)} still fails because each
agent's reads span the entire shared state: Agent~B reads \texttt{geo}'s image while Agent~A is
mid-repair---exactly the stale-read symptom \S\ref{the-agent-paradigm-we-target} warned of.
\textbf{Consistency (C)} then fails as a downstream effect: the final state breaks a global
invariant spanning both agents' writes---``the canary mirrors live
\texttt{geo}''---because B's write to \texttt{geo-canary} was premised on an obsolete value of
\texttt{geo}'s image. Partitioning constrains writes; the bug is in reads.

\textbf{Fork and merge.} A third strategy gives each agent its own copy of the state and merges the
copies once each agent finishes, either via physical workspace forks reconciled by Git, human
reviewers, or another LLM{\cite{cursor2025worktrees,cognition2024devin,replitagent4}}, or via algebraic merges using
predefined state reducers~\cite{langgraph2024}. Three problems limit this approach. First, the merge
itself is hard. A line-level merge (e.g., \texttt{git merge}) catches textual conflicts
but nothing beyond---the merged code can fail to even compile, much less preserve each agent's
original intent. LLM-based review can in principle catch what git misses, but typically fails: the
reviewer must reconstruct each authoring agent's intent (e.g., why a specific workaround was used)
from the merged code alone, and usually cannot. Second, algebraic merges have narrow applicability.
The CRDT family~\cite{shapiro2011crdts} only handles operations that commute; most business logic
does not, and invariants spanning multiple state items lie outside the model entirely. Third, much
of the live state agents interact with---production databases, active Kubernetes clusters---cannot
be forked at all, making the approach inapplicable.

\section{The Goal, and the Gap of Classic Concurrency Control}\label{goal-and-gap}

\subsection{Ideal Goal: Serializability}\label{ideal-goal}

The correctness property this paper targets is what classical transaction theory has long called
\emph{serializability}~\cite{eswaran1976notions,bernstein1987concurrency}: a concurrent run of
several peer agents (not a parent's subagents) over shared state is correct iff its outcome is
equivalent to some serial order of those same agents, in two senses. \textbf{(i)} each agent's
perceived view at every read matches what it would have seen at its serial position;
\textbf{(ii)} the final state of every object matches what that serial order would have
produced.

Applied to the settings of \S\ref{multi-agent-consistency-problem}: a coding session that fires
several agents in parallel should leave the repository as if the requests had run one after
another, and two agents repairing independent cluster failures should leave the cluster as if a
single operator had handled both, sequentially. The canary anomaly violates both senses:
Agent~B's reading of \texttt{geo}'s image is not what a serial execution would have shown it
(violating (i)), and the final state, its canary on the bad image, is not the serial outcome
(violating (ii)).

\subsection{A Recap of Classical Concurrency Control}\label{recap-classical}

Classical concurrency control distinguishes three operation dependencies between transactions
on shared objects~\cite{bernstein1987concurrency}: a \emph{read-from} (wr) where one
transaction reads a value another wrote; a \emph{write-order} (ww) where two transactions
write the same object; and an \emph{antidependency} (rw) where one transaction writes an
object another already read. Reasoning directly about serializability is hard, as it
quantifies over all serial orders and read-from chains, so classical CC narrows to a
graph-checkable subclass: a schedule is \emph{conflict-serializable} iff its
\emph{precedence graph}---a node per transaction, an edge for each dependency---is acyclic,
and conflict serializability is strictly stronger than serializability, hence sufficient. Two
classical protocols enforce that acyclicity in opposite ways.

\textbf{2PL (two-phase locking)}: each transaction takes a shared lock before reading and an
exclusive lock before writing, holds all locks until commit; conflicting operations block, and
the resulting schedule is conflict-serializable by lock-point order~\cite{eswaran1976notions}.
\textbf{OCC (optimistic concurrency control)}: transactions run without locks---reads track
the live state into a read set and writes go to a private buffer; at commit the runtime
validates that no read has been overwritten by a concurrently committed transaction, then
atomically installs the buffer or aborts and retries the whole
transaction~\cite{kung1981optimistic}.

Because an agent's externally visible execution is still a sequence of read and write events
on shared objects, both protocols remain correct in the agent setting. What does not carry
over are the two assumptions classical CC silently makes: that retry is cheap
(\S\ref{performance-gap}) and that the framework offers fork-and-merge semantics
(\S\ref{functionality-gap}).

\subsection{Performance Gap}\label{performance-gap}

\textbf{Long inference makes classical CC cost-prohibitive.} Agent transactions are dominated by
LLM inference and last seconds to minutes---orders of magnitude longer than a
millisecond-scale database transaction.
Classical CC was designed under the assumption that transactions and retries are both cheap;
in agentic scenarios, neither holds.

\textbf{Long lock hold (2PL).} Locks that held for milliseconds in databases now hold for
minutes. While one agent infers for several minutes after reading a shared object, every other
agent waiting on that object blocks. The lock-wait-graph fills up quickly, and deadlocks become
routine rather than exceptional.

\textbf{Costly re-execution (OCC).} A failed validation triggers a full re-execution---all
prior LLM inference, all prior tool calls, and all intermediate artifacts are thrown away. OCC
was designed for cheap transactions where redo is cheap; neither premise holds for agents.

\textbf{Broad read sets.} A third factor amplifies the previous two. LLM agents read aggressively:
they grep entire repositories, list whole namespaces, fetch surrounding context, so the read set is
broad. Under 2PL the locked-object set explodes; under OCC the validated-object set explodes; in
both cases the failure probability rises with the read-set size.

The three factors together turn the agent-CC trade-off sharply against classical protocols.
{In our benchmarks under contention, 2PL deadlocks fire at 0.81 per trial
and its mean speedup over serial execution degrades to 1.04$\times$; OCC
aborts at 0.95 per trial, falls below serial speed (0.93$\times$), and pays
1.83$\times$ token cost on retries (\S\ref{sec:eval-e2e}).}

\subsection{Functionality Gap}\label{functionality-gap}

\textbf{Live state cannot be forked and merged.} Most classical concurrency-control protocols
reach serializability by keeping an in-progress transaction's writes separable from the
published state---either by staging writes in a private buffer and installing them at
commit (OCC, snapshot isolation, Spanner~\cite{corbett2013spanner}), or by writing in place
and rolling them back on abort (2PL during deadlock recovery). Both routes assume the
framework offers \textbf{fork-and-merge semantics}: an uncommitted write can be installed into
the shared state at commit, or rolled back out of it on abort. OCC's commit path needs install
(its abort path is cheap---the buffer was never published, so the framework merely discards
it); 2PL's abort path needs rollback (writes have already gone live, so the framework must
undo them). Either way the framework must support at least one of these two state-manipulation
operations.

For many agentic workloads this precondition does not hold: the framework offers no
fork-and-merge semantics, so neither OCC's install nor 2PL's rollback is implementable. Take
the operations setting of \S\ref{multi-agent-consistency-problem}: a Kubernetes cluster under
debugging is a single live system behind a unique URL---it cannot be meaningfully forked, and
``installing a buffered deployment into a running cluster'' or ``rolling back a deployed
process that has already had side effects'' cannot even be defined: side effects are not
reversible by the framework alone. The same applies to production databases with direct
writes, third-party APIs with side effects, and any other live framework: each is a single
live system whose fork and merge are not part of the contract.

The two gaps push toward two requirements on any agentic CC protocol. \textbf{(R1) Optimistic:}
neither pessimistic blocking nor whole-task retry is affordable (\S\ref{performance-gap}).
\textbf{(R2) Writes take effect immediately:} live state offers no private buffer or fork
(\S\ref{functionality-gap}). \S\ref{insights} shows how agent capabilities make both
realizable, and what mechanism does the reconciliation.

\section{Insights and Our Approach}\label{insights}

\subsection{Notify, Do Not Lock or Abort}\label{observation}

What enables a concurrency control beyond lock and abort is the agent itself: an LLM-powered
agent has three capabilities that classical CC's transaction model cannot exploit.
\textbf{(i)}\label{insight-false-conflicts}~Faced with the broad read sets of
\S\ref{performance-gap}, the agent can judge whether a conflicting write actually invalidates
the premises of its plan. Many such writes do not: a peer appending a one-line log entry to
\texttt{CLAUDE.md} touches an object the agent read but disturbs no premise it relies on.
Classical CC cannot make this distinction and blocks or aborts either way.
\textbf{(ii)}\label{insight-healer}~On a real conflict, the agent does not need whole-task
redo: of the operations it just generated, only the few that depended on the now-stale value
need rewriting, and the LLM can pinpoint them. When Agent~B of
\S\ref{multi-agent-consistency-problem} learns that the repair has replaced
\texttt{geo}'s image, it need not redo the canary task: only the canary's image
field rests on the stale read, and one \texttt{set\_image} repairs it, keeping
the rest of the deployment unchanged.
\textbf{(iii)}\label{insight-undoable}~Even on a live framework with no fork-and-merge
semantics, the agent can produce an inverse action for almost every external write it
issues---a saga-style undo generated from the same context as the original
write~\cite{chang2025sagallm}---so writes that took effect immediately can be reconciled after
the fact. {On Kubernetes, a \texttt{kubectl scale} is inverted by scaling
back to the prior replica count, and a \texttt{kubectl apply} by re-applying
the manifest it displaced (\S\ref{sec:impl-threephase}).}
Capabilities (i) and (ii) make the optimistic posture required by \textbf{(R1)} affordable,
and capability (iii) makes the in-place writes required by \textbf{(R2)} recoverable.

\textbf{Insight.} Classical CC, pessimistic or optimistic, is \emph{mandatory}: the runtime
imposes the remedy for every conflict, by blocking the conflicting operation (2PL) or by
aborting and retrying the transaction (OCC). It has no alternative, because a classical
transaction cannot heal itself: application code can neither judge whether a conflict matters
nor repair its own effects, so the only remedies that work from outside are to delay it or to
discard it. The three capabilities above are precisely what classical CC lacked: a limited
but real form of \emph{self-healing}. The agent can dismiss a conflict that invalidates no
premise (i), regenerate the few operations that a real conflict taints (ii), and undo writes
that should not have landed (iii). A concurrency control that exploits self-healing must
therefore be \emph{advisory} rather than mandatory: the runtime only informs the affected
agent that a conflict may exist, and the remedy is delegated to the agent itself. The
mechanism that carries the advice is \textbf{notification}: it replaces the lock and the
abort as the load-bearing primitive.

\subsection{Overview: Our Approach}\label{approach-overview}

The insight translates into an optimistic execution discipline; \S\ref{detailed-design}
develops it into a protocol with a serializability proof, and \S\ref{implementation} into a
framework. 
The framework fixes a serial order over the
agents at launch and tracks each toolcall's declared read/write footprint. 
Every read is \emph{filtered} (\S\ref{protocol-mtpo}): 
it returns the value the reader's serial position
should see, so far as the writes issued by then allow 
(writes of agents ordered before the reader are visible, writes of agents ordered after are screened out). 
Whether that value is final cannot be known at read time, 
since an agent ordered earlier may still write the same object; 
the agent does not wait to find out: it adopts the value as a premise and proceeds
with its task, optimistically.

Such optimistic execution is reconciled afterwards through two repair channels, split by who decides.
When a write lands that may invalidate a premise some agent has already adopted, the
framework appends a \emph{notification} to that agent's context describing what changed; in
its next inference the LLM judges whether the change is real for its plan
and, when it is, issues corrective reads and writes that patch exactly the affected
operations, or a no-op when nothing it relies on is disturbed. 
When a write has landed in the wrong serial position (a write of lower serial order arrives after it), the speculation itself has misfired; no semantic
judgment is needed, and the framework acts alone:
it invokes the saga-style inverses the
writes registered, unwinds the misordered writes, and reapplies them in
serial order. 

This sketch leaves three design questions.
\textbf{Q1. When and to whom should the framework deliver a notification?}
(\S\ref{protocol-mtpo}).
\textbf{Q2. How should an agent process a notification once it arrives?} 
(\S\ref{protocol-mtpo}).
\textbf{Q3. How must the framework cooperate?} 
(\S\ref{implementation}).

\section{Detailed Design: the Protocol}\label{detailed-design}

The approach of \S\ref{approach-overview} left three questions---when the framework notifies, how
an agent responds, and how the framework cooperates.
We answer them with one protocol, \mtpo{} (Monotonic Trajectory Pre-Order), on a rigorous
model: \S\ref{ideal-goal} named the correctness goal in prose, and stating \mtpo{}'s
guarantee---let alone proving it---requires mathematical precision.

\subsection{System Model and Correctness}\label{system-assumptions}

\phead{Objects and operations.}
The model partitions the shared external state of \S\ref{the-agent-paradigm-we-target} into
\emph{objects}, the units over which reads, writes, and conflicts are defined: a file, a
database row, a Kubernetes deployment, or an abstract grouping of these
(\S\ref{sec:impl-objects}); \(\mathcal{O}\) denotes the object set.
What the protocol must know of each tool call \(\tau\) is its \emph{footprint}: the objects
it reads, \(R(\tau)\), and writes, \(W(\tau)\).
The framework cannot derive the footprint from the call itself---a call's effects can exceed
the arguments it names (a \texttt{kubectl apply} ripples through a reconcile loop), and a
general program does not split into a read set and a write set the way a SQL statement
does---so the footprint enters the system by \emph{declaration}: whoever turns an intent
into a registered tool states its footprint alongside (\S\ref{sec:impl-onboarding}), each
write carrying its class, \emph{blind} or \emph{RMW}
(\S\ref{the-agent-paradigm-we-target}), and, when reversible, an inverse
(\S\ref{sec:impl-threephase}).
We call this the \emph{latent object schema} assumption.
The framework totally orders the tool calls it mediates by \emph{physical execution time},
written \(\prec_t\), and lets each take effect atomically; two calls
\emph{conflict on footprint} when one's write set intersects the other's footprint.

\phead{Two orders, and notification.}\label{notification-mechanism}
Two total orders must stay distinct: the \emph{commit order} \(\prec_\sigma\), the serial
order each agent's run is judged equivalent to, and the physical execution time \(\prec_t\)
at which its tool calls actually reach the framework.
\(\prec_\sigma\) needs a carrier of its own: per object, the \emph{write trajectory}
\(T(o)\) lists the writes on \(o\) in \(\sigma\) order, and its
\emph{materialization}---applying each write, in that order, to \(o\)'s initial state (a
true composition: an RMW write's effect depends on the value before it,
\S\ref{the-agent-paradigm-we-target})---is exactly what the \(\prec_\sigma\)-serial
execution would leave on \(o\).
Call \(T(o)\) \emph{\(\sigma\)-monotone} when its writes arrive in \(\sigma\) order: the
live copy then equals the materialization by construction, and nothing needs repair.
But on a single copy the write overlay cannot be buffered (\S\ref{functionality-gap})---a
write lands in place the moment it executes---and nothing forces arrivals to respect
\(\sigma\).
To repair the reads this breaks, the framework fires a \emph{notification}---a read-only tool
call \(\nu_{i\to j}(o)\) that hands agent \(j\) a value written by agent \(i\), occupying a
\(\prec_t\) position and updating only \(j\)'s read view.

\phead{Agent assumptions.}
We assume three things of the agent, \textbf{A1}--\textbf{A3}, in increasing novelty.
\textbf{A1 (individual success)}: run alone from a state satisfying its premises, the agent
terminates with its intent met.
This is transaction theory's standing premise of individually correct
programs~\cite{bernstein1987concurrency}, with success in place of correctness for a
stochastic executor.
A1's precondition is owed by the task layer: the launched tasks must be \emph{well-posed},
every serial order leaving each agent a premise-satisfying start; else a failure is one of
specification, not of concurrency.
\textbf{A2 (well-formedness)}: every action on shared state goes through a registered tool,
inside its declared footprint, and a delivered notification is consumed before the agent's
next action.
The agent thus follows the protocol, generalizing the \emph{well-formed transaction} of
locking theory~\cite{eswaran1976notions}.
\textbf{A3 (self-healing)}: a notified agent heals correctly.
It recognizes whether the footprint conflict is a \emph{premise conflict}, one whose
changed value invalidates a premise its plan relies on, and rewrites exactly the pending
operations that depended on it.
A2 alone drives the serialization-graph argument; A3 enters where notification perturbs a
run (next); \S\ref{evaluation} measures both.

\phead{Notified serializability.}
Because a notification perturbs an agent's run, we certify correctness through two equivalence
relations.
\emph{Notified equivalence} (\(\cong_N\)): a run with notifications produces the same
observable behavior as the interleaving in which every read directly returned its
\(\prec_\sigma\)-correct value---the notification mechanism merely \emph{realizes} that
interleaving on a stale-prone single copy.
The relation exists only because of A3: self-healing is precisely the ability to fold a
late-arriving value into the plan as if it had been read in place.
\emph{Serialization equivalence} (\(\cong_S\)): an interleaving is equivalent to a serial run,
the textbook relation.
\textbf{Notified serializability} composes the two: the run is \(\cong_N\) an interleaving
that is \(\cong_S\) a serial order in \(\prec_\sigma\).

\subsection{Notification Alone Is Not Enough}\label{notif-serializability}

Notification by itself does not yield serializability.
The protocol it most naturally suggests is \emph{unconditional broadcast}: every agent
writes the single live copy directly, the framework notifies every agent whose footprint the
write touches, and each notified agent heals, rewriting whatever the changed value
invalidates (A3).
This is write-update snoopy cache coherence~\cite{hennessy2019architecture} transplanted
to agents.

However, unconditional broadcast can run into a \emph{livelock}, worse than a
serializability violation (Figure~\ref{fig:livelock}).
The cause is a missing \emph{precedence}.
The framework orders neither agent before the other, so notifications flow in both
directions.
The two antidependency (rw) edges then close a cycle in the precedence graph, ruling out
every conflict-serializable schedule; worse, each agent's view is overturned by the
other's next write.
The classical mechanisms that would impose a precedence are all unavailable on a single live
copy: there is no buffer to stage writes, no fork to isolate readers, and no room for
locks that an agent would hold across minutes of inference (\S\ref{performance-gap}).
Without a precedence, no agent ever holds a stable view to serve as its premises
(\S\ref{protocol-mtpo}).

\begin{figure}[t]
\begin{lstlisting}[aboveskip=2pt, belowskip=2pt, framesep=2pt]
# initial: x = 1, y = 1
# A1's task: x ← y/2        A2's task: y ← x/2
# serial outcomes: (x,y) = (0.5, 0.25) or (0.25, 0.5)
t1  A1: read y = 1
t2  A2: read x = 1
t3  A1: write x = 0.5       → notify A2 (x changed)
t4  A2: write y = 0.5       → notify A1 (y changed)
t5  A1: heal: write x = 0.25  → notify A2
t6  A2: heal: write y = 0.25  → notify A1
t7  A1: heal: write x = 0.125 → notify A2
    ...                       # x → 0, y → 0: livelock
\end{lstlisting}
\caption{\small Unconditional broadcast on the task pair \(x \leftarrow y/2 \parallel
y \leftarrow x/2\). Every write invalidates the other agent's premise and every heal
issues a new write: the run never terminates, although either serial order would
succeed.}
\label{fig:livelock}
\end{figure}

\subsection{Our Protocol: Monotonic Trajectory Pre-Order}\label{protocol-mtpo}

\begin{figure}[t]
\centering
\includegraphics[width=\columnwidth]{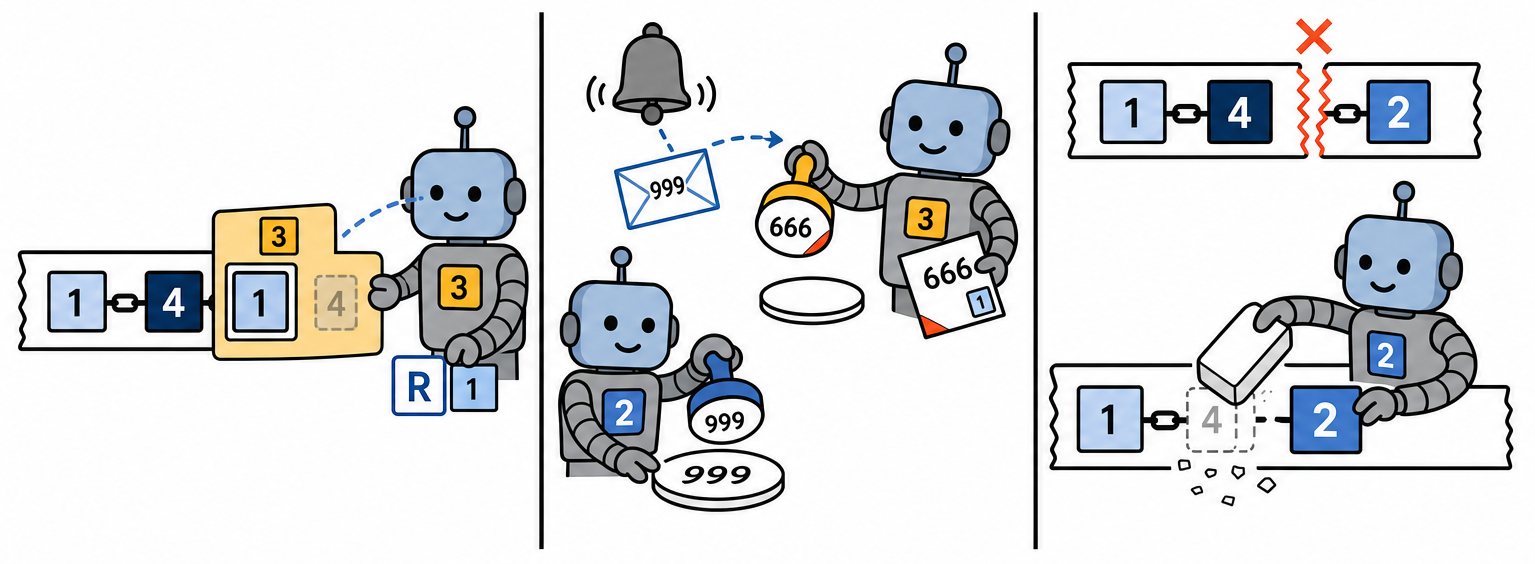}
\caption{\mtpo{}'s three actions: a filtered read returns only lower-\(\sigma\)
values (left); a subscribed notification reports a newer legal value, and the
reader decides whether to repair (middle); a write that would break an
object's \(\sigma\) order first undoes the misapplied writes, then applies
(right).}
\label{fig:mtpo-actions}
\end{figure}

\phead{Key idea: pre-order.}
Supplying the missing precedence constructs the stable view as premise, and it 
is fixed before agent execution: \(\prec_\sigma\) becomes a \emph{pre-order},
assigning each agent a rank \(\sigma_j\) at launch rather than at commit.
A fixed precedence orients every conflict.
Notifications flow strictly from low \(\sigma\) to high, so the dependency graph is a
\(\sigma\)-monotone DAG and no cycle can form.
Each agent thereby holds the stable view the broadcast design lacked:
the state of all writers with \(\sigma<\sigma_j\), which no
higher-\(\sigma\) agent can disturb; absent any notification, that view is already the
framework state at \(j\)'s \(\prec_\sigma\)-serial start, supplying the equivalence
\(\cong_N\) builds on.
On the example of Figure~\ref{fig:livelock}, the lower rank halves once; the higher rank,
notified, halves the new value once more and stops: the serial outcome, reached
concurrently.
What remains is the machinery.
Classical MVTO would keep one version slot per writer and let reads pick by
timestamp~\cite{reed1978mvto,bernstein1983multiversion}; but a slot is a value, so that
machinery silently assumes every write is blind.
RMW writes (\S\ref{the-agent-paradigm-we-target}) force the version store to compose: the
write trajectory of \S\ref{notification-mechanism} replaces the slots, and inserting a
version into it becomes a physical act.
The two mechanisms give the protocol its name, \mtpo{}; it maintains one invariant through
three rules, stated next.

\phead{The protocol.}
An agent is \emph{quiescent on} an object \(o\) when it has no further write on \(o\) to
issue, and \emph{quiescent} when this holds for every object and all delivered
notifications are consumed (a later notification may re-open it).
\emph{GlobalQuiet} is the run state in which every agent is quiescent and no notification
is in flight.

\textbf{\mtpo{}'s invariant}: \emph{at GlobalQuiet}, every object's live copy equals the
materialization of its trajectory, i.e., live state and \(\prec_\sigma\)-serial state
coincide.
Figure~\ref{fig:mtpo-actions} illustrates the three rules.

\textbf{Reads pull from the trajectory (wr).}
A \emph{filtered read} is a synchronous pull:
\(\mathrm{FilteredRead}(j,o)=M(o,\sigma_j)\), the value left on \(o\) by the
entries of \(T(o)\) with \(\sigma\le\sigma_j\).
When those entries end in a blind write, this is simply the \(\sigma\)-greatest version at
or below \(\sigma_j\); an RMW entry, however, composes with everything before it, so in
general the whole prefix must be materialized.

\textbf{Writes apply speculatively (ww).}
R2 (\S\ref{functionality-gap}) admits no buffer, so a write takes effect in place at its
\(\prec_t\) position and joins \(T(o)\) at its rank \(\sigma_j\).
Since nothing forces the two orders to agree, a write \(w\) may be \emph{late}: some entry
\(w'\) of \(T(o)\) satisfies \(w'\prec_t w\) but \(w\prec_\sigma w'\).
Three mechanisms make a late write take effect at its \(\sigma\) rank, not at its arrival
rank, and with it the invariant holds.
If a blind entry above \(\sigma_j\) shadows \(w\), the framework records \(w\) in the
trajectory and never replays it onto the live copy, similar to the Thomas write
rule~\cite{thomas1979majority}.
Readers ranked between the two are served from the trajectory.
Otherwise the framework applies \(w\) after undoing the suffix above \(\sigma_j\) in descending
\(\sigma\), using the inverse each write carries (\S\ref{sec:impl-threephase}).
Last, a write whose tool admits no inverse is held until every agent with
\(\sigma<\sigma_j\) has committed (\S\ref{sec:impl-threephase}): it never speculates,
hence is never late.

\textbf{Notifications push to readers (rw).}
When a lower-\(\sigma\) writer touches an object some higher-\(\sigma\) agent already
read, the framework delivers \(\nu_{i\to j}(o)\), carrying the refreshed
\(M(o,\sigma_j)\), at the receiver's next inference.
A quiescent receiver is re-opened.
\(A_j\) may keep, rewrite, or retract its pending operations (A3).
Classical MVTO would abort the reader here; \mtpo{} instead refreshes it, trading the
abort for a reorder~\cite{mu2014rococo}.
The pull/push division mirrors the two dependencies of \S\ref{recap-classical}.
A read-from (wr) edge needs a lower-\(\sigma\) value, and only the pull can select one: no
push can stop an agent from seeing the too-new value already on the live copy.
An antidependency (rw) edge runs the other way, from lower-\(\sigma\) writer to
higher-\(\sigma\) reader, exactly the direction the push travels.
A two-way notification would reintroduce the cycle of \S\ref{notif-serializability}.

\phead{Correctness.}
We sketch a proof that the three rules deliver notified serializability by constructing the
interleaving that \(\cong_N\) refers to and then showing its precedence graph acyclic.

\textbf{Proposition} (notified serializability).
\emph{Assume A1--A3 and the \mtpo{} rules.
At GlobalQuiet, the run is notified-serializable in \(\prec_\sigma\) order, and every
object's final state is what the \(\prec_\sigma\)-serial execution leaves.}

\textbf{Proof sketch.}
\emph{Step 1 (construct the schedule by \(\cong_N\)).}
A precedence graph is defined over a schedule~\cite{bernstein1987concurrency}, but a run
under \mtpo{} is not a schedule: the framework undoes writes to keep each trajectory
\(\sigma\)-monotone and notifies agents to refresh their views, while a schedule contains
agent operations only.
Let \(I\) be the interleaving in which every read and write takes effect exactly once, at its
\(\sigma\) rank, and every read returns \(M(o,\sigma_j)\) of the trajectory at GlobalQuiet.
The run is \(\cong_N\)-equivalent to \(I\) in two ways.
Writes match exactly by the invariant, since the framework removes the
\(\sigma\)-backward ww edges.
Reads match because the notified agent rewrites exactly the operations that depended on the
changed value, behaving as if it had read the value in place, which resolves the
\(\sigma\)-backward rw edges.
\emph{Step 2 (conflict-serializable by \(\cong_S\)).}
In \(I\) every operation sits at its agent's \(\sigma\) rank, so every wr, ww, and rw edge
of the precedence graph points \(\sigma\)-forward and the graph is a DAG: \(I\) is
conflict-serializable, with \(\sigma\) itself the serial
order~\cite{bernstein1987concurrency}.

\section{The Framework}\label{implementation}

We realize \mtpo{} as \sysfull{}, a single layer of \textbf{toolcall middleware} between the agent framework and the target system.
Every Worker toolcall passes through it: the middleware injects the caller's pending notifications, dispatches the call, and returns the result with any raised during dispatch; notifications can ride on returns because the protocol requires only arrival before the receiver's next inference (\S\ref{notification-mechanism}).
The middleware ships as a plug-in and requires three capabilities from the host, each matching one protocol need: custom tool registration (to interpose), a commit hook (to hold commit until pending notifications drain, the GlobalQuiet premise of \S\ref{protocol-mtpo}), and an A2A channel (\S\ref{sec:impl-onboarding}); all three are standard in LangGraph, OpenAI Agents, Claude Code, and the MCP/A2A specifications~\cite{langgraph2024,openai2025agentssdk,anthropicclaude2025}.

The section follows what the rules consume: the objects reads and writes range over (\S\ref{sec:impl-objects}), the toolcalls held to their declared footprints(\S\ref{sec:impl-objects}), the three routes that serve a read in $\sigma$ order (\S\ref{sec:impl-reads}), the structure that makes writes undoable (\S\ref{sec:impl-threephase}), and the usability problem the discipline creates (\S\ref{sec:impl-onboarding}).

\subsection{The Object Tree and Constrained Tools}\label{sec:impl-objects}

The protocol's rules range over objects, so the system must first fix what an object is.
Objects are organized as a tree, containing two kinds of nodes:
\emph{Natural objects} are units the target system already names: a file, a directory.
\emph{Abstract objects} are units the agent reasons about but no single artifact embodies: a Spark instance, a cluster.
Nodes instantiate lazily, on first mention by a toolcall, keep a stable identity for the session, and carry the object's write trajectory (\S\ref{protocol-mtpo}): its writes in $\sigma$ order.

Every toolcall declares the objects it reads and writes: each tool registers as a structured call whose designated parameters carry object ids, so the middleware reads $R(\tau)$ and $W(\tau)$ off the invocation.
The declaration must also hold at execution; enforcement follows the operation class.
The Worker never writes the request itself: it fills the named slots of a \emph{structured header}, a fixed envelope per tool, and the framework assembles the wire payload, binding each slot to its declared object id.

\subsection{Reads in \texorpdfstring{$\sigma$}{sigma} Order: Snapshot or Undo}\label{sec:impl-reads}

\mtpo{} requires each read to return the $\sigma$-correct value, the materialization of the trajectory prefix at or below the reader's $\sigma$ (\S\ref{protocol-mtpo}); the live copy does not honor $\sigma$, since a higher-$\sigma$ write may land before a lower-$\sigma$ read arrives.
The middleware serves such a read by the cheapest of three routes that applies.

\textbf{Replay on a materialization.}
When state is cheap to copy (files, key-value records, versioned documents), the middleware retains a materialization after every write, and a late read executes against the $\sigma$-legal copy: an analysis script runs over the historical files as if no later write existed, and the higher-$\sigma$ writer is never touched.

\textbf{Record the result.}
Some state cannot be copied, but its read results can: a running container list has no file to copy, so the middleware records the \texttt{docker ps} output after every write under that object, and a late \texttt{docker ps} returns the last $\sigma$-legal recording.
The cost is one read per write; the higher-$\sigma$ container is never rolled back.

\textbf{Live access, backed by undo.}
What remains must run live: an arbitrary SQL query can only run on the database itself (copying it after every write is prohibitive), and a read--modify--write must compose with the writes before it (\S\ref{the-agent-paradigm-we-target}), since an increment computed on a historical copy does not transplant onto the current state.
For both, the middleware brings the live object to the caller's $\sigma$ position: it undoes, in descending $\sigma$, every write above that position, lets the call execute, and notifies each undone agent, which decides whether to retry (\S\ref{protocol-mtpo}).
This route presupposes that every write can be undone; the next subsection makes that hold.

\subsection{Undoability: Three-Phase Toolcalls}\label{sec:impl-threephase}

The undo of \S\ref{sec:impl-reads} needs an inverse for every write it may revoke, and synthesizing one after the fact is often impossible: the write destroys the information the inverse needs, and nothing reconstructs \texttt{DROP DATABASE foo} once it has run.
Undoability is therefore established when the tool is built, before any conflict can arise.
Every tool is written in three segments: \texttt{prepare} runs immediately before \texttt{exec} in a framework-allocated tmp directory and captures everything the inverse will need (for the \texttt{DROP}: a dump of schema and rows); \texttt{exec} carries the tool's intent and modifies the target system; \texttt{reverse} is a self-contained script the framework may invoke at any time to restore the pre-\texttt{exec} state.
The tmp directory is cleared at the owning session's commit, since commit means no further undo.

A few operations admit no \texttt{reverse} at any cost: sending external mail, executing a payment.
Such a tool is tagged \texttt{unrecoverable} at registration; the middleware refuses its calls until every lower-$\sigma$ agent has committed, then pushes an unlock notification.
Concurrency is lost only where no cheaper resolution exists.

\subsection{Generality: The ToolSmith}\label{sec:impl-onboarding}

\begin{figure}[t]
\centering
\includegraphics[width=\columnwidth]{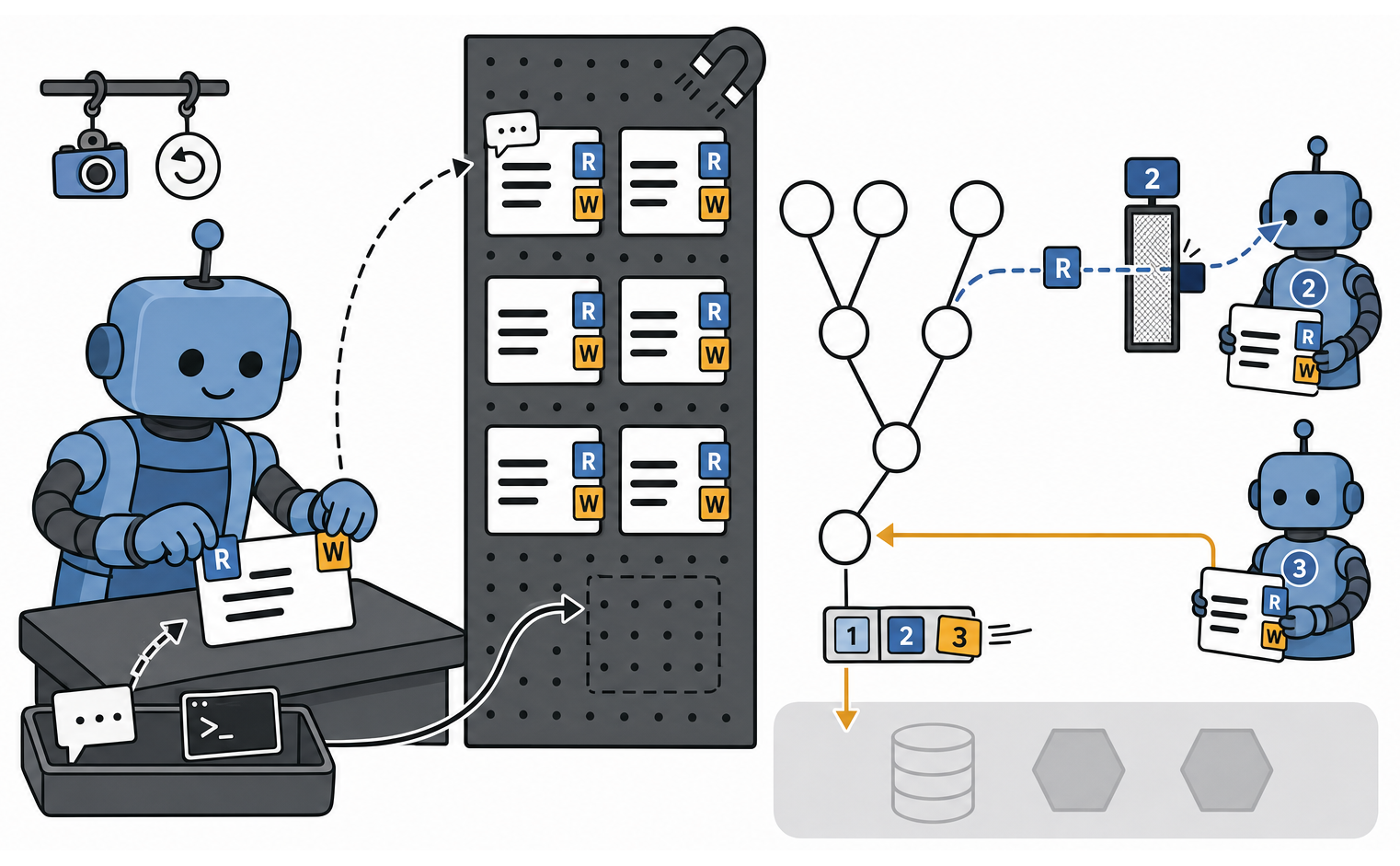}
\caption{\small{On request, the ToolSmith builds binding-carrying tools on the fly;
Worker reads and writes go only through them, under framework control.}}
\label{fig:toolsmith}
\end{figure}

The discipline above creates a usability problem.
Agents are effortless to deploy because one \texttt{bash} covers most of the computing world; \texttt{bash} tracks no read or write set, so this system cannot admit it, and reimplementing every action by hand as a declared three-phase toolcall would cost more than the concurrency it buys.
Nor can the Worker build its own tools: exploration happens precisely where no registered tool exists yet, so those accesses carry no footprint; an untracked read plants a premise the protocol cannot protect, an untracked write changes state without notifying its readers.

The way out is an asymmetry the protocol itself supplies: every conflict in \S\ref{detailed-design} is caused by a write, so an agent that never writes the target system invalidates no premise and needs no concurrency control at all.
The framework therefore hands tool supply to a privileged tool-building agent, the \emph{\sys{} ToolSmith}, unconstrained in reading the target system and forbidden to mutate it.

The ToolSmith works in two phases (Figure~\ref{fig:toolsmith}): on first contact with a new target system, a bootstrap skill drives discovery (list CRDs, namespaces, and verbs on Kubernetes; fetch the schema and probe the RPC surface on MCP) and seeds the object tree and a base tool set from templates (a declarative interface accepts definitions supplied ahead of time); it then stays resident, co-running with the Workers.
When a Worker hits a need no registered tool covers, it submits a synthesis request over A2A, in natural language or, more precisely, as the \texttt{bash} command it wants to run.
The ToolSmith synthesizes from the description or audits the command: it marks the read and write sets, registers objects the tree lacks, attaches \texttt{prepare} and \texttt{reverse}, and returns a constrained tool, complete or as a template whose hole the Worker fills.
The Worker keeps \texttt{bash}-level reach, yet every write it issues flows through a footprint-bound tool.
The ToolSmith's context carries every registered tool, so similar requests deduplicate to an existing one; at steady state most requests hit the table and the overhead amortizes toward zero.
The latent object schema of \S\ref{system-assumptions} is grown exactly this way: seeded at first contact, extended per request at runtime.

\section{Evaluation}\label{evaluation}

This section answers three questions.
\textbf{(Q1)} Does \mtpo{} resolve the concurrent-state-access problem at a cost-performance point better than 2PL and OCC?
\textbf{(Q2)} What does a concurrency anomaly look like under a real run, and how do 2PL, OCC, and \mtpo{} each respond to it?
\textbf{(Q3)} When a target system ships no tool table at all, can the framework grow one online, and at what cost?

\subsection{Setup}\label{sec:eval-setup}

\phead{Environment.}
All experiments run on one dual-socket Intel Xeon E5-2650~v4 server (24 physical cores / 48 threads, 2.20\,GHz; 125\,GiB DDR4; 1.8\,TB NVMe SSD; Ubuntu~22.04).
The K8s scenarios run against a single-node \texttt{kind} cluster on the same host.

\phead{Models.}
Worker agents use \texttt{deepseek-v4-flash} through the OpenRouter endpoint.
A token-rate audit watches for throttling, reconnects, and queueing; any trial that trips one of these is discarded and rerun, so reported wall-clocks are not polluted by supply-side jitter.

\phead{Benchmarks.}
\textbf{WorkBench}~{\cite{workbench2024}} is a simulated office automation harness over five domains (CRM, calendar, email, analytics, project management), driving an agent through natural-language business tasks.
\textbf{AIOpsLab}~{\cite{aiopslab2025}}, from Microsoft Research, supplies live microservice incident investigation and remediation tasks (port misconfiguration, image rollback, permission revocation, and the like) and operates a real K8s cluster through \texttt{bash}.
Neither benchmark was designed for concurrent agents: tasks in each suite are mutually independent and never race for the same object.
We therefore pick five tasks from each suite as the \emph{agent-1} workload and hand-construct a matching \emph{agent-2} for each, chosen so the pair exhibits a textbook concurrency anomaly.
Each cell ships with a set of hand-written \emph{invariants}: any concurrent execution must be equivalent to either \(\textit{agent-1}\cdot\textit{agent-2}\) or \(\textit{agent-2}\cdot\textit{agent-1}\); a run satisfying neither is recorded as a violation.

\phead{Protocols.}
Five protocols are tested on the same middleware (\S\ref{implementation}).
\textbf{serial} runs the two agents back-to-back and sets the correctness and wall-clock-upper-bound reference.
\textbf{naive} runs them concurrently with no coordination, the ``probably correct'' lower bound.
\textbf{2PL} reuses the declarative bindings to acquire read and write locks; a deadlock detector picks a victim and the saga \texttt{reverse} of \S\ref{sec:impl-threephase} unwinds it, with the victim's context cleared before it restarts.
\textbf{OCC} reuses the same bindings under eager validation; the first rw/ww conflict commits the trigger and aborts the conflicting agent, which restarts.
\textbf{\mtpo{}} is the protocol of \S\ref{detailed-design}--\S\ref{implementation}.

\phead{Metrics.}
For every (cell, protocol) pair we run \(N{=}10\) trials and report correctness (the fraction of trials satisfying the invariant), wall-clock mean (relative to serial), and total token-cost mean (relative to serial).
2PL and OCC can livelock, so we cap retries at five: any trial whose victim is aborted five times in a row is terminated and counted as a correctness failure.

\subsection{End-to-End Results}\label{sec:eval-e2e}

\begin{figure*}[t]
\centering
\includegraphics[width=\textwidth]{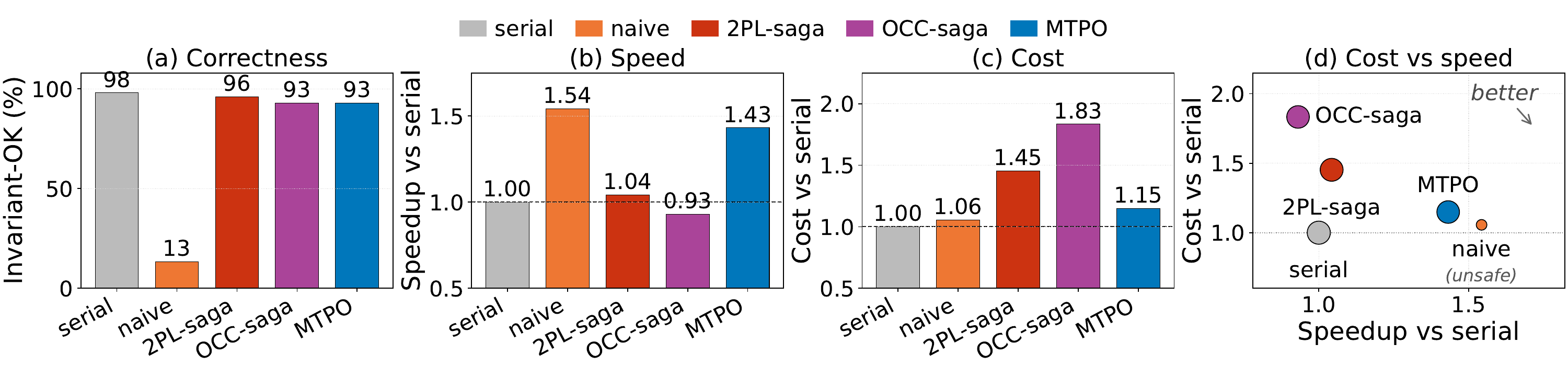}
\caption{\small{Five protocols on ten contended cells: correctness, speedup, token cost, and Pareto front, all normalized to serial.}}\label{fig:eval-page}
\end{figure*}

Figure~\ref{fig:eval-page} reports the five protocols on the ten cells across four panels: correctness, speedup over serial, token cost over serial, and the cost-versus-speedup Pareto front.

\phead{Serial and naive.}
Serial is the cost and correctness optimum: the model spends no attention on coordination and no tokens reasoning about other agents.
Naive is the wall-clock floor; it ignores every dependency and runs the two tasks in parallel.
The achieved speedup sits around \(1.5\times\) rather than \(2\times\) because per-task latency varies, and a pair's wall-clock is bounded by the slower task.
Naive's token cost stays near serial (neither agent does extra reasoning), but on these contended cells its correctness collapses, as expected of a protocol with no coordination at all.

\phead{2PL and OCC.}
Both lift correctness sharply but recover almost no speedup over serial, and add substantial token cost.
Trace analysis shows why.
Under 2PL, deadlocks fire at a rate of \(0.81\) per trial: the victim's progress is reversed and re-driven, collapsing the run back to serial-like latency.
The cause is structural: agents read broadly at start-up and take many shared read locks; the moment either tries to upgrade to a write lock the other holds, the cycle closes.
OCC fares worse: abort rate \(0.95\) per trial, mean token cost \(1.83\times\) serial, peaking at \(2.33\times\) on the worst case, each abort billing the discarded work a second time.
This is the classical heavy-contention regime where pessimistic locking edges optimistic out; agent workloads, with their read amplification (\S\ref{performance-gap}), sit firmly in that regime.

\phead{\mtpo{}}
\mtpo{} clears every cell at a correctness, wall-clock, and cost point materially better than 2PL and OCC.
The pre-order rules out 2PL deadlock and OCC livelock by construction (\S\ref{protocol-mtpo}): notifications travel only low-to-high, the dependency graph is a \(\sigma\)-monotone DAG, and no two agents can wait on each other.
Notification then preserves work that the conflict does not actually invalidate: a notified reader rewrites only the affected fragment of its plan (\S\ref{insight-healer}), and semantically benign syntactic conflicts pass without disturbing the LLM (\S\ref{insight-false-conflicts}).
Both effects show up in token cost, which sits within noise of serial.
Correctness is \(\approx 5\%\) below serial: in five of one hundred trials the notification was delivered but the receiver misjudged its relevance to its own task.
This residual is an A3 (self-healing) gap (\S\ref{system-assumptions}), independent of the protocol; we expect it to shrink with stronger frontier models or targeted finetuning, and leave that to future work.

\subsection{Case Study}\label{sec:eval-casestudy}

\begin{figure}[t]
\centering
\includegraphics[width=\columnwidth]{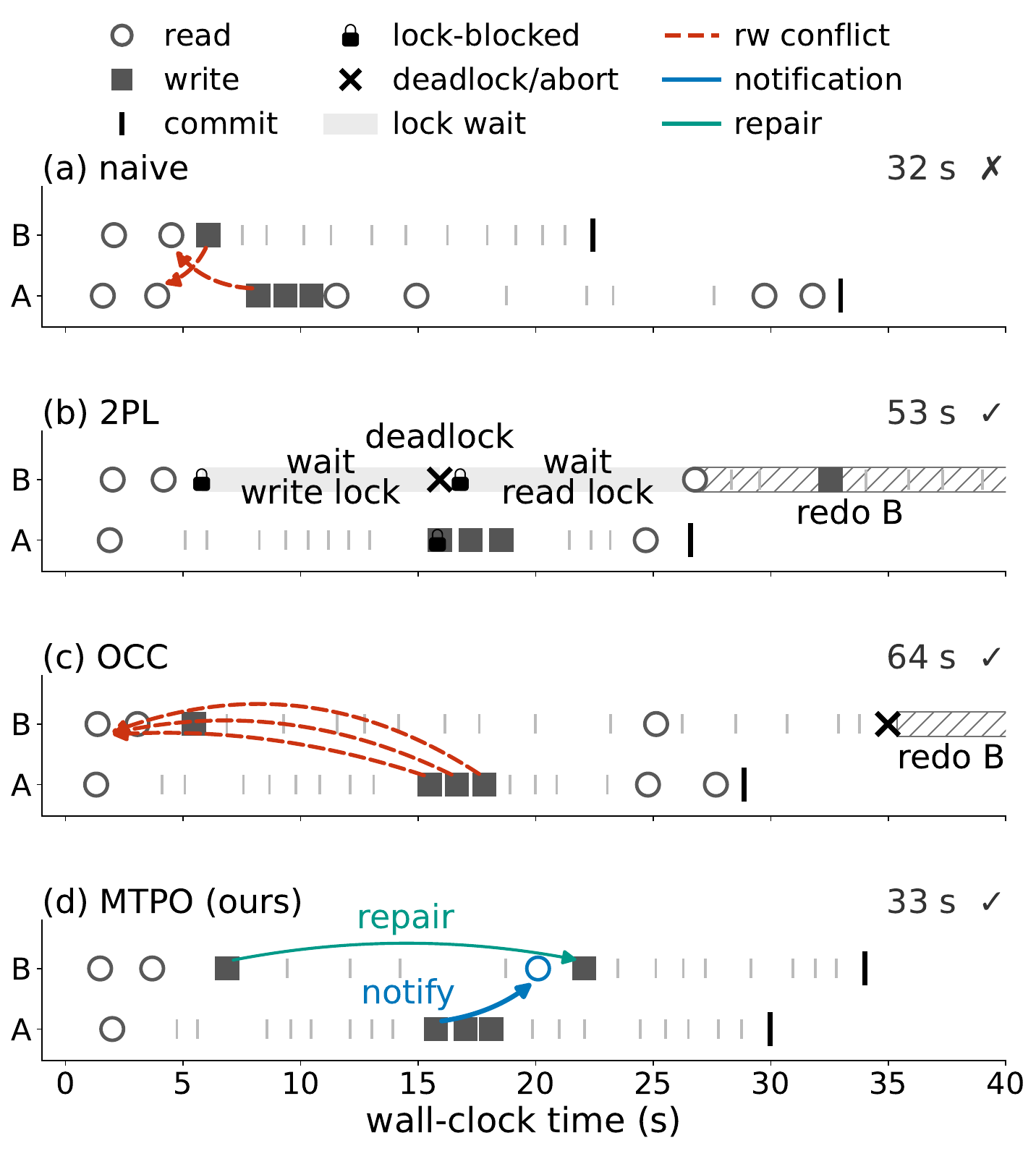}
\caption{\small{The same contended cell under four protocols.}}
\label{fig:casestudy}
\end{figure}

To answer Q2, Figure~\ref{fig:casestudy} takes one of the ten cells and plots the same workload, run under each protocol, on a shared time axis.
The cell is the canary case of \S\ref{multi-agent-consistency-problem}: Agent~A (the repair) and Agent~B (the canary) couple on \texttt{geo}'s image, which B reads as a build premise while A overwrites it, and B's new canary in turn falls within A's audit scope.
The serial baseline is 50.2\,s, and the invariant is the common end state of both serial orders: the canary ends on the canonical image, fully configured.

\textbf{naive (31.8\,s, \xmark).}
This run is the trace of Figure~\ref{fig:canary-trace}: B builds the canary from \texttt{geo}'s pre-fix image, while A's range scan precedes the canary and misses it.
The two rw antidependencies run in opposite directions and cross into a cycle (the two red edges), leaving a non-serializable end state.
Every step is correct at the instant it executes: each read returns the then-current state, and each write's premise still holds when issued.
No stronger model would act differently; the fault sits in the interleaving itself, the same anomaly databases solve with concurrency control.
naive is the fastest precisely because it omits coordination, deferring the cost to the release window.

\textbf{2PL (53.1\,s, \cmark).}
The same two antidependencies surface as lock waits.
Both agents read broadly at start-up and hold shared read locks; B's write-lock request for the new canary falls inside A's range read lock, while A's upgrade of \texttt{geo} from a read to a write lock is blocked by B's read lock.
The wait cycle closes at $t\approx16$\,s: a deadlock.
The detector selects B as the victim: B aborts and releases all its locks, A's pending write lock is granted at that same instant, and A writes \texttt{geo} and its two other fixes.
B then restarts, and its first read hits A's write lock, now held until commit, and blocks a second time until A commits and releases; B redoes all its work from scratch.
B is suspended for about 21\,s and its first execution is discarded entirely, so its wall-clock matches serial and the concurrency gain is zero.

\textbf{OCC (63.7\,s, \cmark).}
Optimistic execution lets both agents run to completion unimpeded, but B's opening cluster-wide audit pulls the whole namespace into its read set, so each of A's three fixes hits it (the three red edges converge on B's single audit at the left).
A commits first and validates; B then fails validation at commit and aborts in full, though every write it issued was correct when issued.
The abort carries no localizing information, so B can only re-audit, re-read, and rebuild from scratch, discarding about 32\,s of first-pass work.

\textbf{\mtpo{} (32.8\,s, \cmark).}
The first 15\,s are identical to naive.
At $t=15.7$\,s, the moment A's write to \texttt{geo} lands, the runtime follows the rw antidependency to B (an in-flight reader that has read \texttt{geo}'s image) and delivers a one-way notification naming the object and its new value (the blue arrow); neither agent stops.
At its next decision point B re-reads \texttt{geo}, confirms the premise has changed, judges for itself that the only affected element is the canary's image, and repairs it in place with a single \texttt{set\_image} (the green arc).
The canary deployment, its label, and its routing entry all stay valid; nothing is discarded and no time is spent waiting.
From A's triggering write to B's completed repair is 6.4\,s, against about 29\,s for OCC's abort-and-redo.
\mtpo{} runs \(1.5\times\) faster than serial at \(1.20\times\) its token cost.

\subsection{Growing the Tool Table Online}\label{sec:eval-toolgrowth}

\begin{figure}[t]
\centering
\includegraphics[width=\columnwidth]{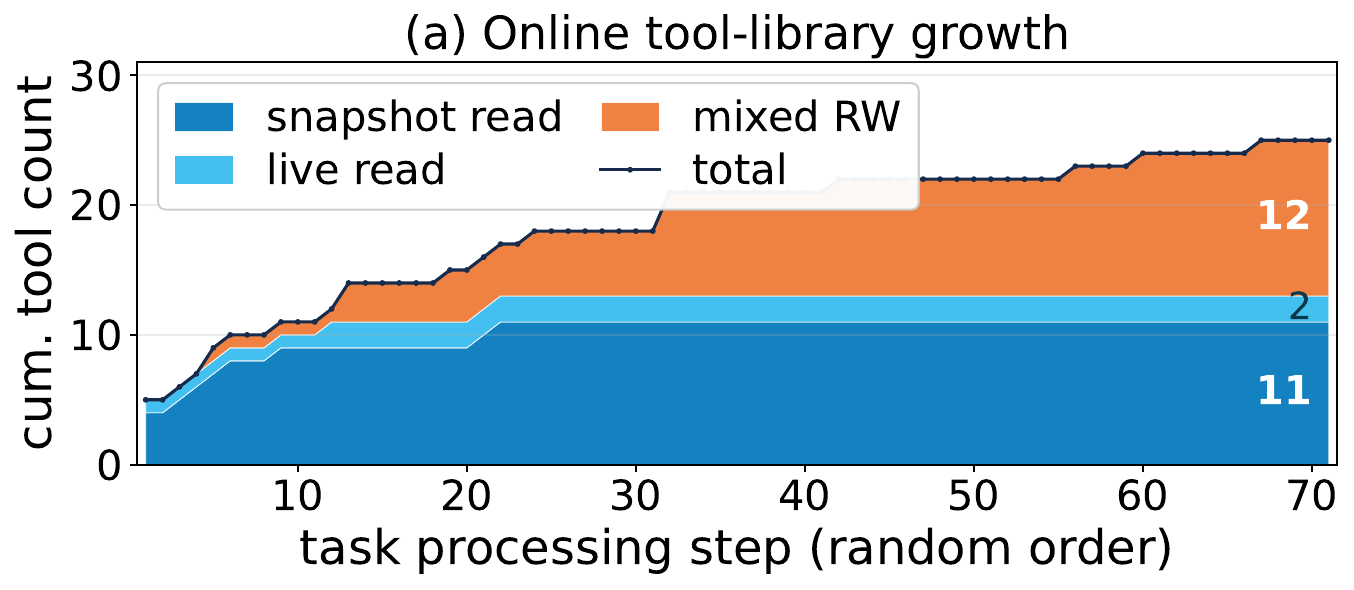}\\[1pt]
\includegraphics[width=\columnwidth]{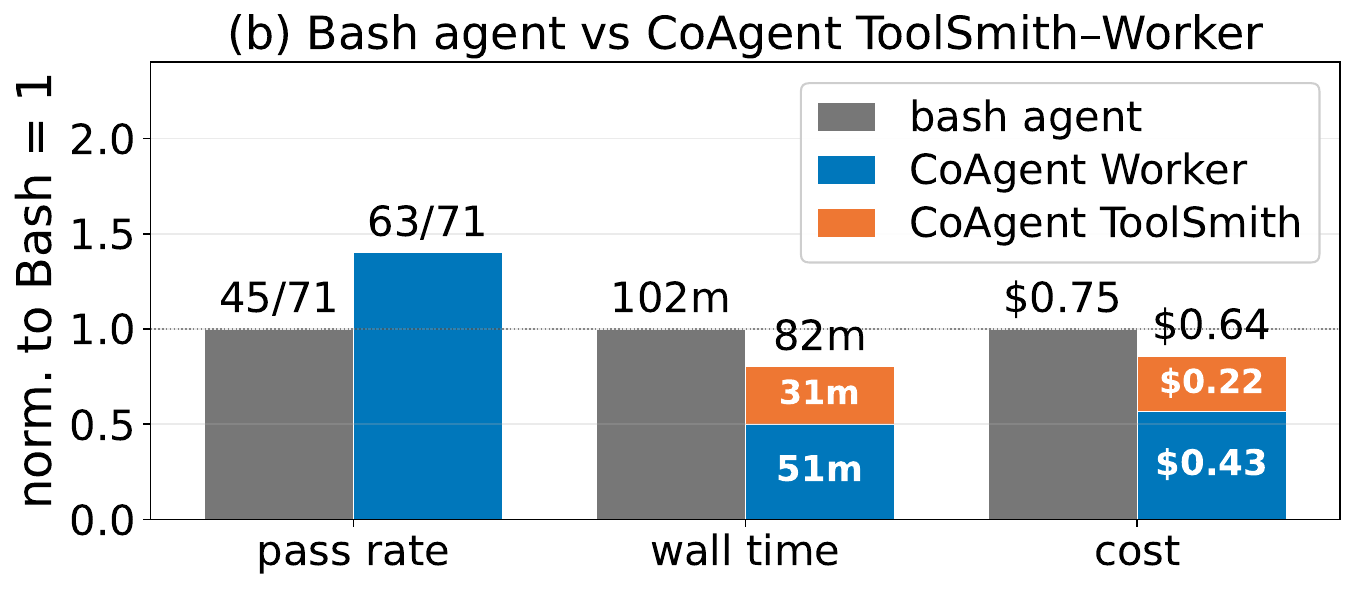}
\caption{\small{(a) Online tool growth; (b) bash agent vs \sys{} ToolSmith--Worker.}}
\label{fig:tool-growth}
\end{figure}

Every comparison so far presupposes a populated tool table.
For WorkBench the presupposition is free: its baseline tools already carry categories and signatures, and a one-time rewrite enrolls them in the registration.
AIOpsLab is the opposite kind of environment: its native agent drives a real K8s cluster with free \texttt{bash}, and no tool table exists at all.
Onboarding it therefore depends entirely on the \sys{} ToolSmith of \S\ref{sec:impl-onboarding}.
This subsection measures that path with one cold-start experiment: whether the table grows on demand, what shape it converges to, and what the ToolSmith--Worker split does to success rate, time, and cost.

\phead{Setup.}
The workload is 71 AIOpsLab tasks in random order, spanning three application stacks (HotelReservation, SocialNetwork, Astronomy Shop) and four task types (detection, localization, root-cause analysis, mitigation); the tool library starts empty.\footnote{AIOpsLab contains 79 problem instances; we exclude eight cases whose fault injection requires Chaos Mesh, which our single-node \texttt{kind} cluster does not run.}
The full suite is harder than the contended cells of \S\ref{sec:eval-setup}: \texttt{v4-flash} solves those, so here both agents run \texttt{deepseek-v4-pro} and every task is capped at 40 rounds.
Before the task starts, a long-lived ToolSmith (one session across all 71 tasks) probes the cluster with read-only \texttt{bash}, registers tools and objects under the prepared-statement discipline, and assigns the task's initial tool list; a Worker that finds its tools insufficient mid-task sends the ToolSmith a natural-language request, and the new tool is hot-inserted at its next step.
The baseline is a pure-\texttt{bash} agent with the same model, task order, and round cap.

\phead{The library converges.}
After 71 tasks the library holds 25 tools: 11 snapshot reads, 2 live reads (logs and events related, which no snapshot can reconstruct), and 12 undoable writes (Figure~\ref{fig:tool-growth}a).
Growth is front-loaded: half the final library exists after 13 tasks (18\% of the stream).
ToolSmith time follows the same curve, from 37\,s per task over the first half to 16\,s over the second, by which point initialization has degenerated into catalog reuse.
All 12 write tools found an executable \texttt{reverse}: \texttt{scale\_deployment}.

\phead{Constraint improves correctness.}
The \sys{} Worker passes 63 of 71 tasks; the \texttt{bash} agent passes 45 (Figure~\ref{fig:tool-growth}b).
Traces attribute the gap to direction rather than capability.
The \texttt{bash} agent has no prior knowledge: it probes the open command space until the cap.
The accumulated tool table provide such prior knowledge of history faults: e.g. the existence of \texttt{list\_service\_ports} suggests comparing ports.
With such a checklist, the Worker resolves the tasks with fewer rounds and with larger success rate.

\phead{Time and cost.}
{\sys} finishes the test within 4910\,s against the baseline's 6118\,s (0.80$\times$).
Log analysis reveals that the primary speedup stems from {\sys}'s structured tool-call list, which enables the agent to complete tasks with fewer rounds.
Cost also lands at \$0.64 against \$0.75 (0.86$\times$), split \$0.43 Worker and \$0.22 ToolSmith.

\section{Related Work}\label{related-work}

\phead{Concurrency control for LLM agents.}
A wave of recent systems brings concurrency control to LLM-agent shared
state. In the taxonomy of \S\ref{observation} all of them are mandatory:
the runtime imposes the remedy, and none proves a correctness property
at agent granularity. One route
tracks read sets at the tool surface and validates at commit:
S-Bus~\cite{khan2026sbus} rebuilds per-agent read sets from HTTP traffic
and machine-checks a partial isolation property, yet measures 74\% of
reads on its SWE-bench workload as invisible to that surface; STORM~\cite{liu2026storm} validates
file-version snapshots at write time. A second route ports one classical
mechanism: conservative 2PL~\cite{calikyilmaz2025optima}, adaptive
locking for SQL agents~\cite{zhou2026atcc},
CRDTs~\cite{pugachev2025codecrdt}, MESI
invalidation~\cite{parakhin2026tokencoherence}.
Atomix~\cite{mohammadi2026atomix} wraps tool calls in progress-aware
transactions at the tool boundary; its measurements reproduce the lock
and OCC cost inversion of \S\ref{performance-gap}. It settles per
resource without proof at single-phase granularity, and leaves
context-carried premises and agent-level serializability out of scope.
SagaLLM~\cite{chang2025sagallm} wraps sequential planning pipelines in
saga compensation and asserts serial-order equivalence for one single
agent; but provide no mechanism or proof enforces multi-agent serializability. 
Our protocol tracks the read view over delivered premises, repairs through
notification, and proves serializability there.

\phead{Classical foundations.}
2PL, OCC, serializability theory, and weak-isolation
definitions~\cite{eswaran1976notions,kung1981optimistic,bernstein1987concurrency,berenson1995critique}
ground this paper; our protocol inherits their provability and changes
the remedy: classical CC blocks or aborts, ours notifies and lets the
agent repair (\S\ref{observation}). \mtpo{} itself descends from
multiversion timestamp
ordering~\cite{reed1978mvto,bernstein1983multiversion} and Calvin's
deterministic pre-ordering~\cite{thomson2012calvin}. Semantic
CC~\cite{garciamolina1983semantic,weihl1988commutativity,li2012redblue}
extracts concurrency from operation semantics but requires static
annotation; our semantic judgment is generated at runtime by the LLM.
The optimistic posture with inverse actions follows
sagas~\cite{sagas1987} and Herlihy's type-specific
compensation~\cite{herlihy1990apologizing}; GoEX~\cite{patil2024goex}
argued for undo in LLM runtimes, and \mtpo{} adds the
\(\sigma\)-monotonicity rule that decides when to compensate.
Collaborative replication~\cite{ellis1989concurrency,terry1995managing,shapiro2011crdts}
shares our structure of a local view, pending operations, and external
updates, but reconciles by algebraic transform or application-supplied
merge; ours reconciles through the LLM, and the broadcast protocol
certifies joint invariants at quiescence, which convergence alone
cannot.

\section{Conclusion}\label{conclusion}

LLM agents increasingly share mutable state, while classical blocking and abort-based concurrency control wastes the long inference work that makes agents expensive.
This paper presents a protocol named \mtpo{}, which handles agents' shared-state access conflicts through targeted repair rather than blocking or aborting, while preserving serializability at quiescence.
Implemented as \sysfull{}, this approach preserves near-serial correctness and token cost while recovering meaningful concurrency on contended workloads.

\bibliographystyle{ACM-Reference-Format}
\bibliography{references}

\end{document}